\documentclass[prl,showpacs,twocolumn,a4paper,amsmath,floatfix]{revtex4}
\usepackage{graphicx}

\begin{document}

\title{Random close packing of polydisperse hard spheres}
\author{Michiel Hermes}
\author{Marjolein Dijkstra}
\affiliation{Soft Condensed Matter group, Debye Institute for
Nanomaterials Science, Utrecht University, Princetonplein 5, 3584
CC Utrecht, The Netherlands } 

\begin{abstract}
We study jammed configurations of hard spheres  as a function of
compression speed using an event-driven molecular dynamics
 algorithm. We find that during the compression, the pressure
follows closely the metastable liquid branch until the system gets
arrested into a glass state  as the relaxation time exceeds the
compression speed. Further compression yields a jammed
configuration that can be regarded as the infinite pressure
configuration of that glass state. Consequently, we find that the
density of jammed  packings varies from $0.638$ to $0.658$ for
polydisperse hard spheres and from $0.635$ to $0.645$ for pure
hard spheres upon decreasing the compression rate. This
demonstrates that the density at which the systems falls out of
equilibrium determines the density at which the system  jams at
infinite pressure. In addition, we give accurate data for the
jamming density as a function of compression rate and size
polydispersity.
\end{abstract}

\maketitle

Dense packings of hard particles are relevant for many
applications ranging from the processing of granular materials to
the development of new materials. In addition, hard spheres have
provided a good starting point in the study of liquids, glasses,
crystals, colloids, and powders. The study of dense packings of
hard spheres has a long history, dating back to Kepler for
crystalline packings and Bernal \cite{bernal1960} for random
packings. In 2005, Hales proved Kepler's conjecture
\cite{hales2005} that the densest packing of identical hard
spheres is achieved by the stacking of close-packed hexagonal
planes yielding a packing fraction $\phi=\pi \sigma^3 N
/6V\approx0.74$. However, when a system of hard spheres is
compressed quickly it does not reach this maximum density and it
jams at lower densities. Jamming phenomena are generic since atomic,
colloidal and granular systems can all be jammed in a state far
out of equilibrium by quickly cooling, compressing or unloading
\cite{liu1998}.

Many authors speculate that the packing fraction of the densest
amorphous hard-sphere configuration is well-defined
\cite{kamien2007,wang2008} although its precise value is not known
as random-close-packing densities $\phi_{\mbox{\small rcp}}$
ranging from 0.634 \cite{wang2008} to 0.648 \cite{speedy1994}
are reported. On the other hand, Torquato {\em et al}
\cite{torquato2000} argued that $\phi_{\mbox{\small rcp}}$ is
ill-defined as the density can always be increased by introducing
crystalline order into the system. Very recently, it is speculated
and/or deduced from mean-field models that jammed configurations 
can be regarded as the infinite pressure limit of glassy states
\cite{parisi,mari}. The reason is that the glass phase for each
density on the metastable fluid branch consists of  a certain
group of glassy states, which will all follow the same metastable
glass branch up to the same jamming density by fast compression
\cite{parisi,mari}.

In this letter, we study jammed configurations of hard spheres as
a function of compression speed. We show that denser amorphous
packings can be generated even without introducing any crystalline
order by lowering the compression speed. However, in the case of
identical spheres, the system (partially) crystallizes for
sufficiently slow compressions.
 Crystallization can be avoided by introducing size
polydispersity. A small size polydispersity affects
$\phi_{\mbox{\small rcp}}$ only slightly but  hampers crystal
growth severely. This allows us to study jamming for a wider range
of compression rates.  We find volume fractions ranging from
$0.638$ till $0.658$ for hard spheres with 10$\%$ size
polydispersity upon decreasing the compression rates.
Additionally, we give accurate data for the jamming density
$\phi_J$ as a function of compression rate and size
polydispersity, which is important for experiments on colloidal
systems.  The packing fraction of a polydisperse colloidal system
is often determined by setting the packing fraction of a
centrifuged sediment \cite{poon} equal to the $\phi_{\mbox{\small
rcp}}$ obtained from simulations
 \cite{schaertl1994,nolan1992}. However, these simulation results are very
 inaccurate and do not take into account any compression rate dependence,
casting doubts on the accuracy of the volume fractions determined
 in experiments via this route. As our results show that the jamming density strongly
depends on the compression rate, we prefer to denote the density
of the jammed configuration  with the jamming density $\phi_J$
rather than with the random-close-packing density
$\phi_{\mbox{\small rcp}}$.

We perform simulations with the event driven molecular dynamics algorithm of
Ref. \cite{lubachevsky1990}. 
Modifications as described by Speedy \cite{speedy1994} were used
to fix the temperature of the system and to define a compression
speed $\Gamma= d\sigma/dt$, where we use the MD time as our unit
of time. In addition the algorithm was adapted to keep the
polydispersity constant during the particle
growth\cite{kansal2002}. The polydispersity was sampled from a
log-normal distribution. The log-normal distribution is nearly
identical to a normal (Gaussian) distribution for small
polydispersities but has the advantage that it is zero for
negative diameters.

\begin{figure}[!hbt]
\includegraphics[width=0.45\textwidth]{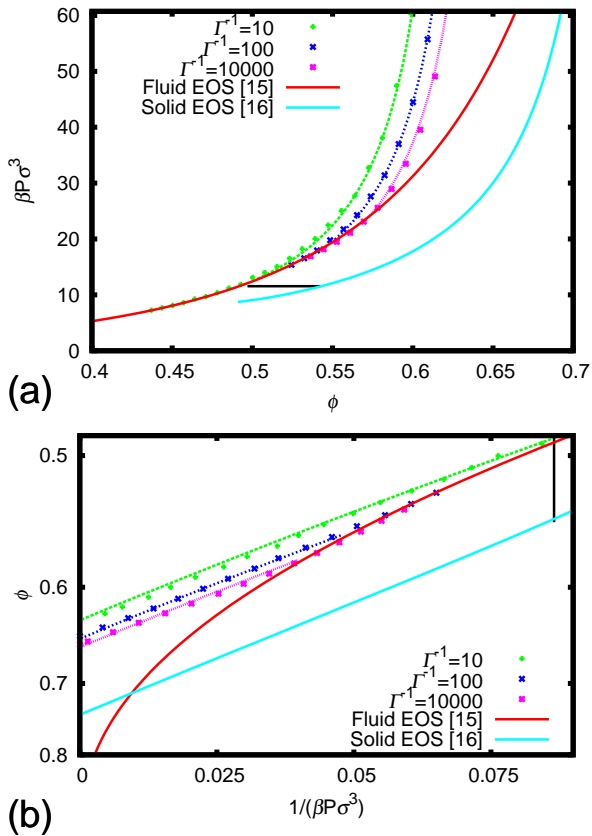}
\caption{ a) Pressure $\beta P \sigma^3$ as a function of packing
fraction $\phi$ for  a system of hard spheres with 10$\%$  size
polydispersity and for varying compression rates $\Gamma$ as
labeled. The red and blue (dark and light full) line denote the
equilibrium equation of state for the fluid and solid phase,
respectively, while the horizontal line denote fluid-solid
coexistence. The dotted lines denote the fits to the simulation data using Eq.\ (1).
b) The same results as in a) but now plotted as a
function of $\phi$ and the inverse pressure $1/(\beta P \sigma^3)$ so that
the infinite pressure limit is clearly visible. } \label{fig_p}
\end{figure}

We determine the equation of state (EOS) from simulations of 2000
particles with a size polydispersity of $10\%$. We average our
results over 50 different runs. To check for finite size effects
we perform  simulations with up to  200000 particles and we find
good agreement within the statistical accuracy.   We plot the EOS
for varying compression rates in Fig. \ref{fig_p}a along with the
equilibrium EOS for the fluid \cite{carnahan1969} and the solid
phase \cite{speedy1998}. We checked for crystallization during our
simulations, but we did not find any crystalline order in any of our
compression runs. We observe that the pressure follows initially
the equilibrium EOS of the fluid phase until the system becomes
arrested into a glass state as the relaxation time of the system
exceeds the compression rate. At this density, the pressure
increases much faster than that of the equilibrium fluid EOS upon
further compression. We find that the density of the jammed
configuration
 at infinite pressure increases with slower compressions. The reason is that
the system has more time to equilibrate for slow compressions, and
hence the system falls out of equilibrium at a higher density on
the metastable fluid branch. Further compression of this glass
phase yields a jammed configuration with a density that is higher
than for fast compressions.  We find clear evidence that the
jamming density of the amorphous packing depends strongly on the
glass transition density, i.e., the density where the system becomes
arrested. Therefore we find a finite range of densities  for the
jamming density depending on the compression procedure and history
of the sample.

This behaviour is well known for molecular glasses
\cite{kauzmann1948}. Indeed, Fig. \ref{fig_p}a resembles the
picture that is found for molecular glasses with $\phi$ and
the inverse of $\beta P\sigma^3$ playing the role of the inverse of the specific
volume and the temperature, respectively. To this end, we plot our
results in Fig. \ref{fig_p}b in the $\phi$ -  $1/(\beta P\sigma^3)$
representation, where $\beta=1/k_BT$ with $k_B$ Boltzmann's constant. We now find
striking similarities with the sketched phase diagrams in Refs.
\cite{parisi,mari}. Recent theoretical calculation using the
replica method by Parisi and Zamponi predict indeed that different
glasses can jam at different densities upon compression, and that
the pressure of the glass phase close to jamming is well-described by a power law
$\beta P/\rho \propto 1/(\phi_J-\phi)$ with $\phi_J$ the jamming
density at infinite pressure\cite{parisi}. 
We observe in Fig. \ref{fig_p}b an almost linear behavior 
for the inverse pressure as a function of $\phi$ for the
glass phase, which we can fit remarkably well over the full 
range using the free volume scaling \cite{kamien2007}: 
\begin{equation}
\beta P\sigma^3=a\frac{\phi_J^{1/3} \phi^{2/3}
}{(\phi_J/\phi)^{1/3}-1},
\end{equation}
where $a$ and $\phi_J$ are fitting parameters. The leading order
term of Eq.\ (1) yields the power law as predicted in [8],
close to jamming.
In addition, the theory predicts an ideal glass
transition at a Kauzmann packing fraction $\phi_K=0.617$ on the
metastable fluid branch, yielding a jammed configuration upon
compression of this ideal glass with a random close packing density of $\phi_{\small
rcp}=0.683$ \cite{parisi}. However, the existence of a
thermodynamic ideal glass transition is still heavily debated. We
note that our results do not depend on the existence or
non-existence of such a glass transition. Our results show a
nonequilibrium glass transition at densities in the volume fraction range of 0.50-0.59,
which is far below the theoretical predictions for the ideal glass. This is
to be expected as the structural relaxation time diverges on
approaching the ideal glass transition \cite{berthier}. Hence, it
is impossible to reach $\phi_K$, since already at lower densities,
the fluid gets arrested in a nonequilbrium glass as the relaxation
time becomes longer than the simulation time.

\begin{figure}[!hbt]
\includegraphics[width=0.45\textwidth]{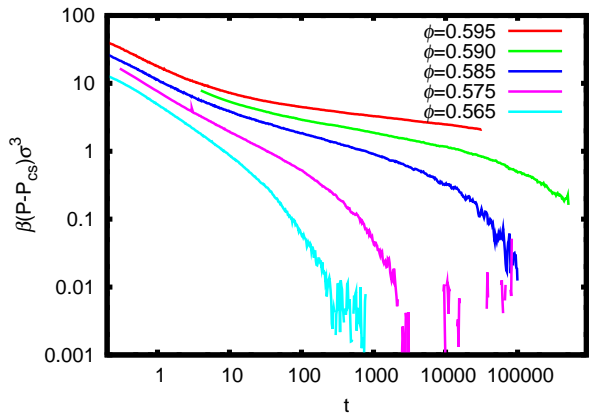}
\caption{The pressure difference $\beta (P-P_{CS})\sigma^3$ with
respect to the Carnahan-Starling equation of state $P_{CS}$
obtained from constant volume simulations as a function of MD time
for a system of hard spheres with 10 $\%$ size polydispersity. The
runs are started with an initial configuration obtained from a
fast compression. }\label{fig_pl}
\end{figure}

In order to investigate the divergence of the structural
relaxation time on approaching $\phi_K$, and to estimate what the
maximum density is at which we can still equilibrate a state on
the metastable fluid branch, we perform constant volume
simulations of a system of 20000 spheres with a 10$\%$ 
size polydispersity and varying packing  fractions. We start the
runs  with an initial configuration obtained from a fast
compression. Fig. \ref{fig_pl} shows the pressure difference with
respect to the Carnahan-Starling EOS as a function of time. We
clearly see that the pressure initially decays towards an
intermediate state for all $\phi$. Subsequently, large collective
rearrangements are required to relax the system further. Finally,
we observe that the pressure reaches the value predicted by the
equilibrium Carnahan-Starling EOS.  The time scale for the system
to equilibrate to the equilibrium fluid phase is comparable to the
relaxation time $\tau_\alpha$ that can be determined from the
decay of the intermediate scattering function  \cite{berthier}. We
clearly observe in Fig. \ref{fig_pl} that the relaxation time
diverges with packing fraction. The  equilibration time of a
system with 20000 particles at $\phi=0.585$, costed more than $10^5$ MD time
steps, which is equivalent to two weeks on a desktop PC. The equilibration
time for $\phi=0.59$ is expected to be more than 20 weeks. Hence, the ideal glass
transition at $\phi_K$, and the corresponding random close
packing that can be achieved by compressing the ideal glass, are
both inaccessible. Instead the system will fall out of equilibrium
into a non-equilibrium glass state at a density that depends
strongly on the compression.

\begin{figure}[!hbt]
\includegraphics[width=0.45\textwidth]{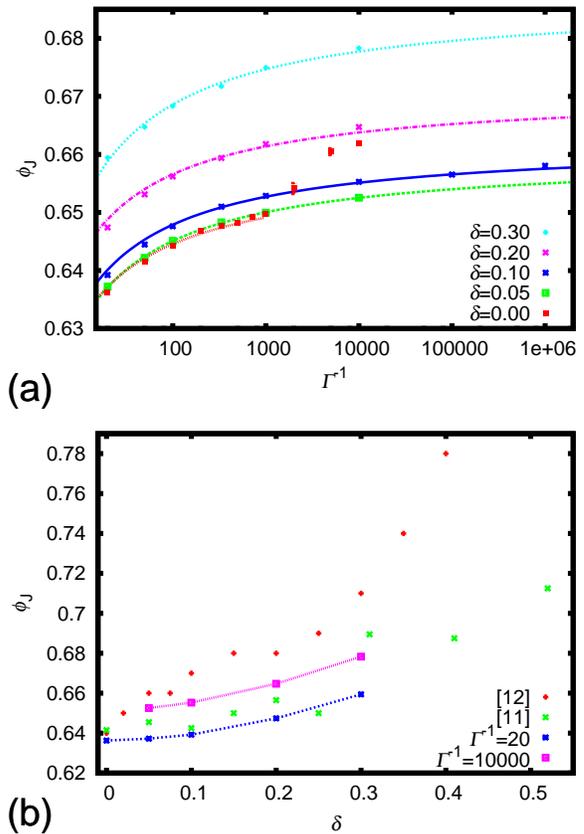}
\caption{a) Jamming density $\phi_J$ as a function of the inverse
compression rate $\Gamma^{-1}$ for polydispersities $\delta$ as reported in
the labels. The lines denote fits to our results (see text). For
monodisperse hard spheres, the line ends where (partial)
crystallization is observed. b) Jamming density  $\phi_J$  for
$\Gamma^{-1}= 20$ and $10^4$ as a function of size polydispersity $\delta$.
The data from Nolan {\em et al.} \cite{nolan1992} and Schaertl
{\em et al.} \cite{schaertl1994} are denoted by the crosses and
plusses, respectively. }\label{fig_cmpr}
\end{figure}

We therefor study systematically the jamming density as a function of compression
speed for several size polydispersities $\delta$. We perform simulations of
2000 particles using varying compression rates and we terminate
the simulations when the time between successive collisions
becomes of the same order of magnitude as our numerical accuracy,
yielding pressures $\beta P\sigma^3$ up to $100000$ for slow
compressions. To determine the jamming density $\phi_J$, we fit
the equation of state close to jamming (the last few volume
percent) with Eq.\ (1). We average our
results over 50 different runs. Fig. \ref{fig_cmpr}a shows the
jamming density as a function of the inverse compression rate for
varying polydispersities $\delta$. We have fitted these extrapolations with
$\phi=a+b/\log(\Gamma)$, where $a$ and $b$ are fitting parameters.
The jamming density $\phi_J$ for pure hard spheres ranges from
$0.635-0.645$ for $10 \leq \Gamma^{-1} \leq 1000$. For faster
compressions $\Gamma^{-1}<10$, the simulations do not yield jammed
configurations and the system (partially) crystallizes for
$\Gamma^{-1}>1000$ as can be observed in Fig. \ref{fig_cmpr} as
$\phi_J$ increases rapidly. We also observe that the jamming
density $\phi_J$ increases with increasing size polydispersity.
For $\delta=10 \%$, we find that $\phi_J$ varies from
$0.638-0.658$. As the size polydispersity prevents
crystallization, we are now able to study $\phi_J$ for compression
rates across five orders of magnitude. Although the slope of the
curves decreases with increasing $\Gamma^{-1}$ (slower
compressions), it is hard to justify an extrapolation to
infinitely slow compression rates. Figure \ref{fig_cmpr}b shows the
jamming density as a function of size polydispersity $\delta$ for a
compression rate $\Gamma^{-1}=20$ and $10^4$. For comparison, we
also plot the data from Nolan \cite{nolan1992} and Schaertl
\cite{schaertl1994}. Our results are close to those of Nolan
\cite{nolan1992}. The results from Schaertl {\em et al.}\ deviate
from our results but  can be explained since they have done single
runs with low accuracy \cite{schaertl1994}. The strong dependence
of the jamming density on the compression speed explains the range
of densities that has been found in the literature,  obtained by
different authors and by using different algorithms. Our results
also show that a size polydispersity of up to $5\%$ does not
increase the jamming density significantly from the monodisperse
case. A much larger dependence on the polydispersity is often used
in the experiments \cite{poon} based on the results of Ref.
\cite{schaertl1994}, casting doubts on the precise values for the
volume fractions determined in experiments via this route.

In conclusion, we have studied the jamming density of hard spheres
as a function of compression speed. We find that during the
compression, the pressure follows closely the metastable liquid
branch until the system gets arrested into a glass state. Further
compression of the glass phase yields an almost linear behavior
for the inverse pressure as  predicted theoretically
\cite{kamien2007,parisi}, providing evidence that the jammed configuration
can be regarded as the infinite pressure configuration of that
glass state. We find, as expected, higher jamming densities for
slower compression. The reason for this is that the system has
more time to equilibrate for slower compressions, and hence the
density increases at which the system falls out of equilibrium.
 Consequently, we find that the density of
jammed packings varies from  $0.635$ to $0.645$ for pure hard
spheres upon decreasing the compression rate. For slower
compressions we observe partial crystallization and for faster
compression our simulations do not result in jammed states. For
slightly polydisperse (10$\%$) hard spheres, the jamming density
varies from $0.638$ to $0.658$. Additionally, we give accurate
data for the jamming density as a function of compression rate and
size polydispersity. Our results provide evidence that
there is a fundamental difference between the ideal glass 
transition \cite{parisi}, which takes place at a fixed density (if it does exists at all),
and the jamming transition \cite{nagel}, that takes place at a range of densities, 
as was already speculated by \cite{parisi,mari}. However, there is
a link between the density at which the systems falls out of
equilibrium and the density at which the system jams at infinite pressure.

\end{document}